\documentclass[a4paper,prl,aps,superscriptaddress,citeautoscript,twocolumn,twoside,showpacs,longbibliography]{revtex4-1}

\usepackage{amsmath}
\usepackage{graphicx}
\usepackage{amssymb}
\usepackage{epsfig}
\usepackage{amsfonts}
\usepackage{color}

\newcommand{\eg}{{{e.g.}}}
\newcommand{\ie}{{{i.e.}}}
\newcommand{\etal}{{\it{et al.}}}
\newcommand{\US}{\mathrm{US}}

\newcommand{\ket}[1]{| #1 \rangle}
\newcommand{\bra}[1]{\langle #1 |}
\newcommand{\oprod}[1]{\ket{#1}\bra{#1}}

\newcommand{\beq}{\begin{eqnarray}}
\newcommand{\eeq}{\end{eqnarray}}

 \def\>{\rangle}
 
\newcommand{\tr}{\mathrm{tr}}
\newcommand{\Tr}{\operatorname{tr}}%from Miguel

\begin{document}

\title{Experimental temporal quantum steering}

\author{Karol Bartkiewicz}
\email{bark@amu.edu.pl} \affiliation{Faculty of Physics, Adam
Mickiewicz University, PL-61-614 Pozna\'n, Poland}
\affiliation{RCPTM, Joint Laboratory of Optics of Palack\'y
University and Institute of Physics of Academy of Sciences of the
Czech Republic, 17. listopadu 12, 772 07 Olomouc, Czech Republic }

\author{Anton\'{i}n \v{C}ernoch}
\affiliation{Institute of Physics of Academy of Science of the
Czech Republic, Joint Laboratory of Optics of Palack\'y University
and Institute of Physics of Academy of Sciences of the Czech
Republic, 17. listopadu 50A, 77207 Olomouc, Czech Republic}

\author{Karel Lemr}
\affiliation{RCPTM, Joint Laboratory of Optics of Palack\'y
University and Institute of Physics of Academy of Sciences of the
Czech Republic, 17. listopadu 12, 772 07 Olomouc, Czech Republic }

\author{Adam Miranowicz}
\affiliation{CEMS, RIKEN, 351-0198 Wako-shi, Japan}
\affiliation{Faculty of Physics, Adam Mickiewicz University,
PL-61-614 Pozna\'n, Poland}

\author{Franco Nori}
\affiliation{CEMS, RIKEN, 351-0198 Wako-shi, Japan}
\affiliation{Department of Physics, The University of Michigan,
Ann Arbor, MI 48109-1040, USA}

\begin{abstract}
Temporal steering  is a form of temporal correlation between the
initial and final state of a quantum system. It is a temporal
analogue of the famous Einstein-Podolsky-Rosen (spatial) steering.
We demonstrate, by measuring the photon polarization, that
temporal steering allows two parties to verify if they have been
interacting with the same particle, even if they have no
information about what happened with the particle in between the
measurements.  This is the first experimental study of temporal
steering. We also performed experimental tests, based on the
violation of temporal steering inequalities, of the security of
two quantum key distribution protocols against individual attacks.
Thus, these results can lead to applications for secure quantum
communications and quantum engineering.
\end{abstract}

% 03.67.Mn Entanglement measures, witnesses, and other characterizations
% 42.50.Dv Quantum state engineering and measurements in quantum optics
% 89.70.+c Information theory and communication theory

\date{\today}
\maketitle

Einstein-Podolsky-Rosen (EPR) steering refers to strong
nonclassical nonlocal bipartite correlations. It was first
described by Schr\"odinger~\cite{Schrodinger35} as a
generalization of the EPR paradox~\cite{EPR35}. The recent
celebration of the 80 years of steering and the EPR
paradox~\cite{EPR80} showed that our understanding of this
phenomenon is now much deeper, but still very limited. Steering
differs from quantum entanglement~\cite{Horodecki09} and Bell
nonlocality~\cite{Bell64,Genovese05,Brunner14}, as not every entangled state
manifests steering and not every state that manifests steering
violates Bell's inequality~\cite{Wiseman07}. In other words,
steerable states are a subset of entangled states and a superset
of Bell nonlocal states. Analogously to Bell nonlocality, steering
can be detected independently of other nonclassical correlations
by simple inequalities~\cite{Wiseman07,Cavalcanti09, Smith12} that
can include as little as two measurements with two outcomes for
Alice and a set of four possible states for
Bob~\cite{Cavalcanti09}. Such inequalities were tested in several
experiments~\cite{Saunders10, Walborn11,
Smith12,Bennet12,Handchen12, Steinlechner13,Su13,Schneeloch13},
including a recent loophole-free experiment~\cite{Wittmann12}.
Steering can be interpreted as a correlation between two systems
(measuring devices), where only one of them is trusted.  This
property shows an operational meaning of steering and indicates
its potential applications in quantum cryptography and quantum
communication, \eg, for entanglement
distribution~\cite{Branciard12,Wiseman07}. Steering-based
protocols can provide secure communications even when only one
party trusts its devices. Such protocols are easier to implement
than completely-device-independent protocols~\cite{Acin07}, but
are more secure than standard protocols requiring mutual trust
between the communicating parties.

Temporal steering~\cite{Chen14} (TS), analogously to EPR steering,
is observed when Alice can steer Bob's state into one of two
orthogonal states by properly choosing her measured observable.
Despite this similarity, the implications of these temporal and
spatial phenomena are fundamentally different. To detect
TS~\cite{Chen14}, Alice and Bob perform consecutive measurements
(using a  random sequence of  mutually-unbiased bases known only
to them) on the same system to test temporal correlations between
its initial and final states. Breaking the temporal steering
inequality, given in Ref.~\cite{Chen14}, implies that no
unauthorised party can gather full information about the final
quantum state. In other words, there was no quantum collapse and
the observed correlations are stronger than any correlations
between the initial state and its classical copy prepared by
measuring and resending the initial state. Such strong temporal
correlations must have a quantum origin. Their stronger form
asserts that  no third party can gather more information about the
original state than Bob.  In this case, Alice and Bob witness
temporal correlations of a unique strength, which prove that they
interact with the original quantum system and not with one of its
quantum copies. This unique relation between the past and the
future is referred to here as monogamous quantum causality. Here,
Alice mostly steers the future state of Bob and no other system
can be steered with the same strength. This regime is especially
interesting for quantum cryptography, because it allows performing
secure quantum key distribution protocols over a quantum channel,
which is not fully characterised (trusted).

In contrast with EPR steering~\cite{Ou92, Hald99,Bowen03,
Howell04,Saunders10, Walborn11,Wittmann12, Bennet12,Handchen12,
Smith12,Steinlechner13,Schneeloch13,Su13,Chen2013,Sun14,Chen2015,Wollmann2016,Sun2016}, TS has not yet
been investigated experimentally. This article reports, to our
knowledge, the first experimental demonstration of TS. We verify,
in a quantum linear-optical experiment, the relation between TS
and two quantum key distribution (QKD) protocols based on mutually
unbiased bases (MUB). Specifically, we apply temporal steering for
experimental testing the security of the Bennett-Brassard 1984
protocol (BB84)~\cite{BB84} and the six-state 1998 protocol by
Bruss (B98)~\cite{Bruss98} against individual attacks. As
discussed theoretically in Ref.~\cite{Bart15}, the
unconditional security of these
protocols~\cite{secBB84,secB98,GisinRMP} (even against individual
attacks) implies the existence of a kind of monogamous temporal
correlations. The first experimental test of this temporal
steering monogamy is reported here.

TS is understood  as  the ability of Alice to prepare a quantum
object in a quantum state that after travelling, for a period of
time through a damping channel, to Bob will manifest strong
temporal correlations between its initial and final states. These
correlations tell us how strong is the influence of Alice's choice
of observable on Bob's results. The channel can erase partially or
completely Alice's influence. This decoherence process will take
some time. Thus, TS is an appropriate name for this effect. It was
shown~\cite{Chen14} that these temporal correlations are related
to the one-way security bound in BB84. Therefore, this new kind of
steering, similarly as the standard EPR steering, can be
responsible for secure (one-way) quantum communications. However,
Ref.~\cite{Chen14} did not explain the origins of this relation
between TS and QKD.
In a certain sense, this TS is a kind of one-way
(or asymmetric) (temporal) steering because of the time arrow. For
spatial steering, one could consider one-way (spatial) steering as
well as two-way (spatial) steering, where the roles of Alice and
Bob are interchanged. For TS this can be done only for a unitary 
(reversible) evolution of a given steered system.
Steering is, by definition,
asymmetric, as corresponding to one-side device-independent
entanglement detection. Thus, in steering one assumes that only
one side is performing faithful measurements. This is in contrast
with entanglement, where both parties are trusted, as well as Bell's
nonlocality, where both parties are untrusted.

%------------------------------------------------------------------
\emph{Experimental temporal steering.---} In our experiment, Alice
with probability $P(a|A_i)$ prepares qubits by rotating $\ket{H}$
(a horizontally-polarised photon)
to one of the six eigenstates of the Pauli operators. This is done
by the consecutive use of half- ($\mathrm{H}$) and 
quarter-wave ($\mathrm{Q}$) plates as shown
in Fig.~\ref{fig:1}. To implement BB84, Alice sends
eigenstates of only the $\sigma_1$ and $\sigma_2$ operators. She
implements B98 by including also $\sigma_3$. 
This method of state preparation is equivalent to performing a projective
nondestructive measurement $A_i$ by separating states of $a=+1$
and $a=-1$ with a polarising beam splitter (an equivalent of the
Stern-Gerlach experiment~\cite{Gerlach22}), and detecting their
presence in one of its paths. A photon with probability
$P(+1|A_i)$ chooses the path designated for $a=+1$ and, with
probability $P(-1|A_i)$, the path for $a=-1$. However, the latter
approach would be much more difficult to implement because it
would require a nondemolition photon-presence detection (see, \eg,
Ref.~\cite{Bula13}). In the former approach, we assume that the
state preparation governed by the probability distribution
$P(a|A)$ is equivalent to Alice's nondestructive equiprobable
measurements of $A_i=\sigma_i$ for $i=1,2,3$, where the outcomes of the
measurement $A_i$ are $a=\pm1$ and appear with the probability
$P(a|A_i)=1/2$. 
The final measurement does not have to be nondestructive, 
because  this is the final step of the measurement.
This is sufficient to check if the channel can
 partially or completely erase Alice's influence on the 
state detected by Bob, hence to demonstrate TS
without applying a quantum  nondemolition measurement.

A nontrivial case of TS requires a nonunitary dynamics. We
implement such evolution with  two channels labelled by $\lambda$.
To show this evolution, we analyze the TS parameters $S_N \equiv \sum_{i=1}^NE\left[ \left\langle
B_{i,t_{\mathrm{B}}}\right\rangle _{A_{i,t_{\mathrm{A}%
}}}^{2}\right]$
corresponding to the left-hand-side of the TS inequality of Chen
\etal~\cite{Chen14} (see also Ref.~\cite{Bart15}) and a
measure of TS, i.e., the so-called TS weight $w_{t,N}$, as defined
in Refs.~\cite{Bart15,Chen16}. 
The TS inequality is satisfied for all classical states and it reads
\begin{equation}\label{eq:ts_ineq}
S_N \equiv \sum_{i=1}^NE\left[ \left\langle
B_{i,t_{\mathrm{B}}}\right\rangle _{A_{i,t_{\mathrm{A}%
}}}^{2}\right] \le 1.
\end{equation} 
The technical definition of   
$w_{t,N}$ is rather complex, thus we present it in the Methods. 
However, this quantity measures the amount of  the genuine temporal steering correlations between Alice and Bob, 
maximised over the possible Bob's measurements. The TS parameter $S_N$ depends on the number $N=2,3$ of unbiased measurements
$B_i\equiv\sigma_i$ performed by Bob. This corresponds to a sum over the measurements of the expectation values
$ E\left[ \left\langle B_{i,t_{\mathrm{B}}}\right\rangle _{A_{i,t_{\mathrm{A}
}}}^{2}\right] \equiv \sum_{a=\pm 1}P(a|A_{i,t_{\mathrm{A}}})\left\langle
B_{i,t_{\mathrm{B}}}\right\rangle _{a|A_{i,t_{\mathrm{A}}}}^{2},$
where Bob's outcomes are related to the state
projection performed by Alice, as
$\left\langle B_{i,t_{\mathrm{B}}}\right\rangle_{a|A_{i,t_{\mathrm{A}
}}}\equiv \sum_{b=\pm 1}b\,P(b|A_{i,t_{\mathrm{A}}}=a,B_{i,t_{\mathrm{B}}}
).$ The parameter $N$ represents the number of the MUB
used by Bob to analyze the received qubit.
For only one channel
(corresponding to a simple unitary evolution of an isolated
photon), the TS parameters and the TS weight would not exhibit an
interesting behaviour beyond simple oscillations. 
In the extreme case, when the output state of a channel is always
the same, independent of the input state, Alice's influence on the
state detected by Bob is completely erased. However, for typical
imperfect channels, this Alice's influence is only partially
lost and, in the context of QKD, this loss of information can be attributed to 
eavesdropping.

For both BB84 and B98 there exists a minimal value of the average quantum bit error rate
(QBER) $r_N$ in the raw key for which the respective protocol is no longer secure. 
For individual qubit attacks these values are
$r_2=\tfrac{1}{2}(1-\tfrac{1}{\sqrt{2}})$ for BB84 ($N=2$) and
$r_3=\tfrac{1}{6}$ for B98 ($N=3$). 
These values correspond to the minimal amount of noise
introduced by an eavesdropper equipped with a
quantum cloning machine optimised to copy the states
prepared by Alice in a relevant protocol. 
The security of these QKD protocols is naturally related
to optimal quantum cloning and it was studied in various works~(see,
\eg, Refs.~\cite{GisinRMP,Soubusta07, Bartuskova07,Lemr12,
Bartkiewicz13prl} and references therein). 
It was recently shown that the TS parameter $S_N$
depends on the average QBER $r_N$ or the average fidelity 
$F_N=1-r_N$ of the states received by Bob\cite{Bart15}.
This was done  by expressing the
steering parameter $S_N$ in terms of the fidelity $F_{i,a}$ of the
particular states prepared in a QKD protocol by Alice with respect to 
the states measured by Bob, i.e., 
$S_N \equiv \frac{1}{2}\sum_{i=1}^N \sum_{a=\pm 1} (2F_{i,a} - 1)^2.$
The derived relation between temporal steering and security of QKD\cite{Bart15}
$S_N>N(1-2r_N)^2$ asserts that $\mathrm{QBER}_N<r_N.$
Thus, the minimal violation of this security condition indicates the maximal values of
$S_N=N(1-2r_{N})^2$ for which the relevant QKD protocols are
insecure. 

In our
experiment, $\lambda$ can have two values 0 and 1. For
$\lambda=0$, a photon of polarisation $V$ is erased, with
probability $p_0=1-\tau$, with a filter of transmittance $\tau$.
For $\lambda=1$, with probability $p_1=1-p_0$, the photon is passed to Bob in the
state $R(\theta)\ket{a|A}$. The polarisation is rotated by an
angle $\theta$, \ie, it is transformed by the operator
$R(\theta)=\openone \cos{\theta} + i \sigma_2\sin{\theta},$ 
where $\openone$ is the two-dimensional identity operator. 
Each time a photon is erased by the filter, Bob counts one photon
received in state $R(\theta)\ket{H}$. We can set the values of
$p_0$ and $p_1$ by setting the transmission rate $\tau$, while {we
set the rotation angle $\theta$ by inserting two half-wave plates
into the beam (see Fig.~\ref{fig:1}). We set in the experiment
$\tau$ and $\theta$ in a way that Bob receives states  which can
be expressed as
\begin{equation}\label{eq:state}
\hat{\rho}_{a|A_i}(t_B) = R(4t_B)\left(\begin{array}{cc}
1-\rho_{11}\mathrm{e}^{-t_B} & \rho_{01}\mathrm{e}^{-\tfrac{t_B}{2}}\\
\rho_{10}\mathrm{e}^{-\tfrac{t_B}{2}} & \rho_{11}\mathrm{e}^{-t_B}
\end{array}
\right)R^\dagger(4t_B),
\end{equation}
where $\rho_{00}\equiv \bra{+1,A_3} \hat{\rho}_{a|A_i}(t_A)
\ket{+1,A_3} $, $\rho_{01}=\rho_{10}^*\equiv \bra{+1,A_3}
\hat{\rho}_{a|A_i}(t_A) \ket{-1,A_3}$, $\rho_{11}\equiv
\bra{-1,A_3} \hat{\rho}_{a|A_i}(t_A) \ket{-1,A_3}$, and
$\hat{\rho}_{a|A_i}(t_A)=\oprod{a,A_i}$. The time $t_B$, between
Alice's and Bob's measurements, is measured in units of the
inverse of the damping constant $\gamma$. In our experiment, we
set $\gamma = 1/t$, where $t=50$\,ns is the time needed for a
photon to make one loop in the setup. The time $t_B=-t\log{\tau}$
is set by changing the value of $\tau$. The states in
Eq.~(\ref{eq:state}) correspond (up to a unitary rotation) to  the
solution of a simple relaxation~\cite{Zagoskin11book} model that
provides an example of nonunitary dynamics.

Finally, Bob performs polarisation analysis with a setup
consisting of a set of a half- and quarter-wave plates, a polarising beam splitter (PBS),
and a single-photon counting module (SPCM). This allows him to project
the incoming photons on each of the $\ket{b,B}$ states. Bob
receives photons arriving from both channels $\lambda$. As a
result of Bob performing his projections on each of the six
states $\ket{b,B_j}$, we obtain the probability distribution
$P(a,b|A_i,B_j)$ (we know what state has been sent by Alice) that
holds the same information as the assemblage
$\lbrace\rho_{a|A_i}\rbrace_{a,i}$. This is because
$P(b|A_i=a,B_j)=\tr{(\ket{b,B_j}\bra{b,B_j}\rho_{a|A_i})}$.
However, in the context of the QKD protocols we are interested
only in the compatible bases ($i = j$), therefore,
$P(b|A_i=a,B_j)=\tr{(\ket{b,B_j}\bra{b,B_j}\rho_{a|A_i})}\delta_{i,j}$,
where $\delta_{i,j}$ is the Kronecker delta.

Figures~\ref{fig:2}, \ref{fig:3},
and~\ref{fig:4} show that our experimental data are
in good agreement with the expected results. However, the
correspondence is not perfect due to experimental imperfections.
From these measured results we calculated the TS parameters
$S_N$~\cite{Chen14,Bart15} and the corresponding TS
weights $w_{t,N}$~\cite{Bart15,Chen16}. The TS inequality
[see Fig.~\ref{fig:2}] provides a sufficient condition for
the existence of TS and a security threshold for the MUB protocols
with symmetric noise against individual attacks. In our experiment
this threshold for $S_N$ is usable only for BB84 ($N=2$) at a
specific time $t_B=n\pi/2$, where $n=0,\,1,\,2$. In these cases,
BB84 is secure against individual attacks if the TS inequality is
violated, \ie, $S_2>1$. However, it is not so for B98, where the
protocol can be insecure even if $S_3>1$. In B98, we deal with the
asymmetric dynamics of the channel (\ie, a relaxation process to one
of the eigenstates of the Pauli operators); however, we can assess
the security using the $S_3>4/3$ condition. The increase for
$t_B\gamma > 0.8 $ of the $S_2$ parameter with respect to the
theoretical curve, excluding the $R(4t_B)$ rotation around the $y$
direction, is caused by interchanging the noise between the $z$
and $x$ directions.  The TS weight, as shown in Fig.~\ref{fig:3}, 
which is insensitive to rotations, proves or disproves the
existence of TS. Comparing Figs.~\ref{fig:2}
and~\ref{fig:3}, it is clear that the relation between the
TS inequality and the TS weight is not trivial. The TS weight
implicitly includes all possible TS inequalities, so it detects
steering better than the TS inequality. In
Fig.~\ref{fig:3}, the value $w_{t,N}=0$, implies the
insecurity of the relevant QKD protocol.

%------------------------------------------------------------------

\section*{Discussion}

We note that a well-known technique for analysing the security in
QKD is to introduce virtual entanglement by conceptually replacing
state preparation with measurements on an entangled
source~\cite{NielsenBook}. Thus, one could think that the standard
steering inequalities applied to the virtual entangled source can
be used to determine the security requirements on the preparation
and measurement correlations. Nevertheless, the described idea
corresponds to analysing the security in QKD via spatial or
spatio-temporal steering. In contrast with this idea, we analysed
the security in QKD via purely temporal steering  by replacing the
two-qubit measurements with measurements on a single qubit,
followed by the evolution under some noisy quantum channel.

In optical fibres one deals with several types of noise and losses
which limit the range of the applicability of QKD. These problems are  the
polarization-dependent losses, geometric phase, birefringence, 
and polarization mode dispersion (see Ref.~\cite{GisinRMP}).
In our experiment, we implemented a combination of 
the polarization-dependent losses and polarization rotation
[see Eq.~(\ref{eq:state})]. 
The polarization-dependent losses can be significant in 
components like phase modulators and open a way for attacking QKD
protocols, e.g., the two-state protocol  \cite{Bennett92} 
by changing nonorthogonal states into orthogonal ones 
\cite{Huttner96}. However, the state-dependent losses are 
usually not that important in optical fibres.
The polarization rotation could be attributed to the geometric 
phase (a special case of the Berry phase \cite{Berry84}) that 
accumulates, e.g., when polarised photons are transmitted 
through fibre loops. Alternatively, this polarization rotation 
can be caused by polarization-dependent dispersion
due to the stress applied to optical fibres. The latter effect occurs
in the polarization controllers used in our setup 
shown in Fig.~\ref{fig:1}.

Our analysis remains unchanged after including
such additional state-independent losses like lossy quantum
channels or imperfect detectors. State-dependent losses 
may cause some basis states to be transmitted more often than others. 
Thus, the security threshold should be revised by
replacing the previously used $r_N$ with its optimal value 
found for the new asymmetric qubit distribution.
This value can be calculated efficiently by optimising the
the average single-copy fidelity $F=1-r_N$ of
1$\rightarrow$2 qubit cloners~\cite{qmoney}.

%------------------------------------------------------------------

\section*{Conclusion}

We experimentally demonstrated the
possibility of temporal quantum steering with photon polarizations
in a linear-optical setup. We applied TS for testing the security
of two popular quantum-key distribution protocols (i.e., BB84 and
B98), which are based on mutually-unbiased bases. We have measured
the evolution of the TS weight~\cite{Bart15,Chen16} and
TS parameters corresponding to the violations of the TS
inequalities of Chen \etal~\cite{Chen14}. 
To our knowledge, this is the first experimental
determination of the TS weight. Note that this TS weight is
closely related to spatial steerable
weights~\cite{Skrzypczyk14,Piani15}, which have not been measured yet.
Our experimental tests
demonstrate the monogamy of TS and, thus, the security of the
analysed cryptographic protocols against individual attacks. We
believe that these first experimental demonstrations of TS can
lead to useful applications in secure quantum communication.

\section*{Methods}

\textbf{Pauli operators in the photon-polarisation basis.}
We apply the standard eigenstate expansions of the Pauli
operators, which read as: $\sigma_1 = \oprod{+}-\oprod{-}$,
$\sigma_2 = \oprod{L}-\oprod{R}$, and $\sigma_3 =
\oprod{H}-\oprod{V}$, where
$\ket{-1,A_1}\equiv\ket{-}=(\ket{H}-\ket{V})/\sqrt{2}$,
$\ket{+1,A_1}\equiv\ket{+}=(\ket{H}+\ket{V})/\sqrt{2}$,
$\ket{-1,A_2}\equiv\ket{R}=(\ket{H}-i\ket{V})/\sqrt{2}$,
$\ket{+1,A_2}\equiv\ket{L}=(\ket{H}+i\ket{V})/\sqrt{2}$,
$\ket{-1,A_3}\equiv\ket{V}$, and $\ket{+1,A_3}\equiv\ket{H}$.
These eigenstates of the Pauli operators are, respectively: the
antidiagonal, diagonal, right-circular, left-circular, vertical,
and horizontal polarization states.\\

\noindent 
\textbf{Experimental setup. }
Alice's setup, as shown in Fig.~\ref{fig:1} (in the main article) consists
of $\mathrm{Q}_{A}$ and $\mathrm{H}_{A}$ that allow her to set any
of the $\ket{a,A}$ states. The polarisation modes are flipped
$V\leftrightarrow H$ by $\mathrm{H}_{1}$, then separated by
$\mathrm{BD}_{1}$ and $H$-polarised photons are attenuated by the
$\mathrm{NDF}$. Next, the polarisation modes are recombined by
first flipping back the polarisation modes $V\leftrightarrow H$ by
$\mathrm{H}_{2}$ and then joining the beams at $\mathrm{BD}_{2}$.
The channel performs the operation $R(\theta)$ with wave plates
$\mathrm{H}_3$ and $\mathrm{H}_{4}$. Each of these two plates
implements  the transformation that flips the polarisation
direction along their optical axes. It can be readily shown that
the two transformations constitute a rotation by angle $\theta  =
2\delta$, where $\delta$ denotes the angle between the optical
axes of the two wave plates. The polarisation controllers are used
to stabilise the output polarisation. To satisfy the consistency
conditions~(see Ref.~\cite{Bart15}) we assume that all the photons
sent by Alice reach Bob, \ie, we interpret all the physical photon
losses due to the imperfections of the photon counting process as
the result of state preparation and not as the true transmission
losses. However, the photons in the state $R(4t_B)\ket{V}$, which
are lost due to the NDF of transmittance $\tau$, are added to the
final counts, \ie, the NDF is interpreted as a part of Bob's
detection setup.

Bob's setup consists of $\mathrm{Q}_{B}$  and $\mathrm{H}_{B}$
followed by a PBS and SPCM (Perkin-Elmer). This setup allows to
project the incoming photons onto every of the states $\ket{b,B}$.
The beam splitter (BS) is used to verify if Alice has indeed
prepared photons in the desired state $\ket{a,A}$ before sending
them through the channel. However, the purity of states sent by
Alice is $p\approx96\%$, \ie, Bob's results are effectively scaled
by the shrinking factor $s=\sqrt{2p-1}\approx 96\%.$ Moreover, the
BS rotates the photons travelling to Bob by circa $7^\circ$ around
the $z$ axis with respect to the photons travelling to Alice. We
take these factors into account in the presented theoretical
curves, unless stated otherwise. The time delay, between the
photons send via the delay loop and reflected to Bob, directly
allows to analyse both the input and output states using the same
detection setup.  Single photons are generated using a heralded
single-photon source. This source uses a type-I spontaneous
parametric down-conversion (SPDC) process occurring in a 1\,mm
thick BBO crystal pumped by the third harmonics (355\,nm) of a
Nd-YAG laser (300\,mW) with a repetition rate of 2500\,Hz and a
pulse width of 6.5\,nm. The signal photon generated in the SPDC
process powers the experiment, while the idler is used for
triggering. We registered circa 2000 such photon pairs per second.
The triggering allows us to post-select only on valid detection
events (by eliminating detector dark counts) and  to gate the
signal detection corresponding to the direct reflection on the
beam splitter BS shown in Fig.~\ref{fig:1} of the article (no runs in the
loop) from one, or possibly more runs, in the fibre loop. 
Our source delivers signal photons with a polarization-state purity of about
$96\%$ and a beam transversal  profile corresponding to the TEM$_{00}$
mode filtered
by single-mode fibers. The generation rate was adjusted so that the
probability of more than one photon impinging on the detector was
limited to about $5\%$.\\

\noindent
\textbf{Experimental losses.} The experimental data collected by Bob are shown in Fig.~\ref{fig:4}. The setup
implements the intended transformations with an average fidelity
of circa $95\%$. We used this value to estimate the sizes of the
average error bars presented in Figs. \ref{fig:2}, \ref{fig:3}, and \ref{fig:4}. Moreover, the
setup dephases the transmitted photons, which results in the
attenuation of the off-diagonal density matrix terms by an
additional factor of $\exp(-0.05t_B)$.
The setup  introduces polarization-dependent losses, which
are described by the  ratio of the maximum achievable transmissivity 
for $H$-polarised photons and the maximum transmissivity for 
$V$-polarised photons (no $V$ polarisation filtering) equal to $T(H)/T(V)= 96\%$.
Finally, there are some technological polarization-independent 
losses that do not affect our results, leading to
a single-loop transmissivity $T_\mathrm{tech} = 10\%$.

\vspace*{5mm}

\noindent
\textbf{Temporal steerable weight}
The temporal steerable weight is a counterpart of
the EPR steering weight of Skrzypczyk \etal~\cite{Skrzypczyk14},
where the assemblage $\{\rho_{a|A_i}(t_A)\}_{a,i}$
is formed by  the set of Alice's measurements and
outcomes. The conditional probability $P(a|A_i)=\Tr[\rho_{a|A_i}(t_A)]$
of Alice detecting the outcome $a$ while setting her apparatus
to measure $A_i$ can be calculated directly from the assemblage. 
Bob at time $t_B$ receives states $\hat{\rho}_{a|A_i}(t_A)= \rho_{a|A_{i,t_{\mathrm{A}}}}(t_A) / P(a|A_{i,t_{\mathrm{A}}})$
after they passed through a nonunitary channel.
Thus, Bob's assemblage is
$\{\rho_{a|A_i}(t_B)\}_{a,i}\equiv\{\rho_{a|A_i}\}_{a,i}$,
where the explicit form of $\rho(t)$ is given in the main article.
To obtain this assemblage experimentally,
Bob performs quantum state tomography of the received qubit.

The unsteerable assemblages~\cite{Skrzypczyk14} 
can be created independently of Alice's observables,
and can be expressed as
\begin{equation} \label{eq:unsteerable}
\begin{aligned}
&\rho_{a|A_i} = \sum_\gamma D_\gamma(a|A_i)\rho_\gamma &\forall a,i, \\
\text{such~that}& \quad\quad \Tr \sum_\gamma \rho_\gamma = 1,
\quad\quad \rho_\gamma \geq 0 &\forall \gamma,
\end{aligned}
\end{equation}
where $\gamma$ is a random variable,
$\rho_\gamma$ are the states received by Bob, and
$D_\gamma(a|A_i)$ are deterministic functions which assign
$\gamma$ to a specific measurement
$A_i$ and its outcome $a$~\cite{Skrzypczyk14,Bart15,Chen16}. 

The TS weight $w_t$ is the minimal amount of strictly
steerable resources needed to split any assemblage as
\begin{equation}\label{eq:sw_def}
    \rho_{a|A_i} = w_t\rho_{a|A_i}^\mathrm{S} + (1-w_t)\rho_{a|A_i}^\US\quad\quad \forall a,i,
\end{equation}
where $\rho_{a|A_i}^\mathrm{S}$ is a steerable assemblage and
$\rho_{a|A_i}^\US$ is an unsteerable assemblage. The smallest possible value 
of $0\leq w_t \leq 1$ in Eq.~(\ref{eq:sw_def}) corresponds to
the TS weight. Thus, to find its value one needs to solve 
a convex optimization problem. For small matrices, this can be done efficiently using semi-definite programming.

%\end{document}

%------------------------------------------------------------------
%\bibliographystyle{naturemag}
%\bibliography{bib_TS}

%------------------------------------------------------------------
\section*{Acknowledgements}

K.L. and K.B.
acknowledge the financial support by the Czech Science Foundation
under the project No. 16-10042Y and the financial support of the
Polish National Science Centre under grant
DEC-2013/11/D/ST2/02638. A.\v{C}. acknowledges financial support
by the Czech Science Foundation under the project No.
P205/12/0382. The authors  also acknowledge the project No. LO1305
of the Ministry of Education, Youth and Sports of the Czech
Republic financing the infrastructure of their workplace. F.N. is
partially supported by the RIKEN iTHES Project, MURI Center for
Dynamic Magneto-Optics, JSPS-RFBR contract no. 12-02-92100,
JST-IMPACT, CREST, and a Grant-in-Aid for Scientific Research (A).
A.M. and F.N. acknowledge the support of a grant from the John Templeton Foundation.

\section*{Author Contributions}

K.B. and A.M. developed the theoretical framework. K.B. planned
the experiment, processed the experimental data, and wrote the
paper. A.\v{C}. and K.L. designed and built the experimental setup
and performed the measurements. All authors discussed the results
and participated in the manuscript preparation.

\section*{Additional information}

\noindent\textbf{Competing Interests:} The authors declare that
they have no competing financial interests.

\noindent\textbf{Correspondence and requests for materials} should
be addressed to K.B.~(email: bark@amu.edu.pl) concerning
theoretical aspects and to A.\v{C}~(email: acernoch@fzu.cz) and
K.L~(email: k.lemr@upol.cz) concerning experimental aspects.

\begin{figure*}[ht]
\centering
\includegraphics[width=0.6\linewidth]{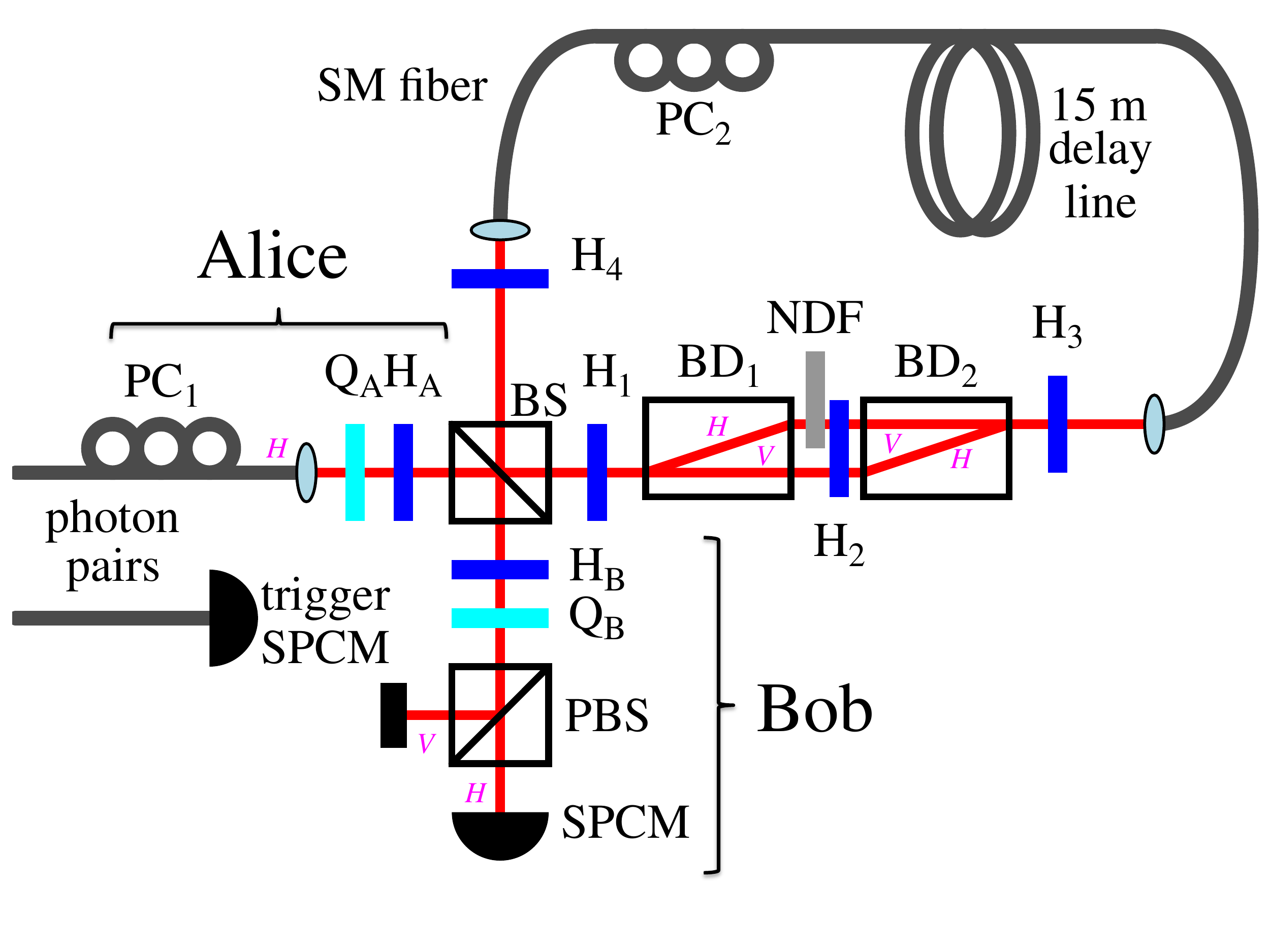}
\caption{Experimental setup for demonstrating
temporal steering. Here we use abbreviations: BS for beam
splitter, PBS for polarising BS, NDF for neutral density filter of
tunable transmittance $\tau$, $\mathrm{BD}_{n}$ for beam dividers,
$\mathrm{PC}_{n}$ for the polarisation controllers compensating
polarisation rotation in the fibers, $\mathrm{Q}_{n}$ for
quarter-wave plates, $\mathrm{H}_{n}$ for half-wave plates, SPCM
for single-photon counting module, SM for the 15\,m long
single-mode optical fiber which forms a delay line of $t=80$\,ns
(this is larger than the $50$\,ns long dead time of the SPCM).
Alice prepares the initial state by rotating a $H$-polarised
photon with $\mathrm{Q}_\mathrm{A}$ and $\mathrm{H}_\mathrm{A}$
plates. Bob performs state analysis by setting his measurement
basis with $\mathrm{H}_\mathrm{B}$ and $\mathrm{Q}_\mathrm{B}$,
and detecting the incoming photon in one of the orthogonal
polarisation states. }
\label{fig:1}
\end{figure*}

\begin{figure*}[ht]
\centering
\includegraphics[width=0.45\linewidth]{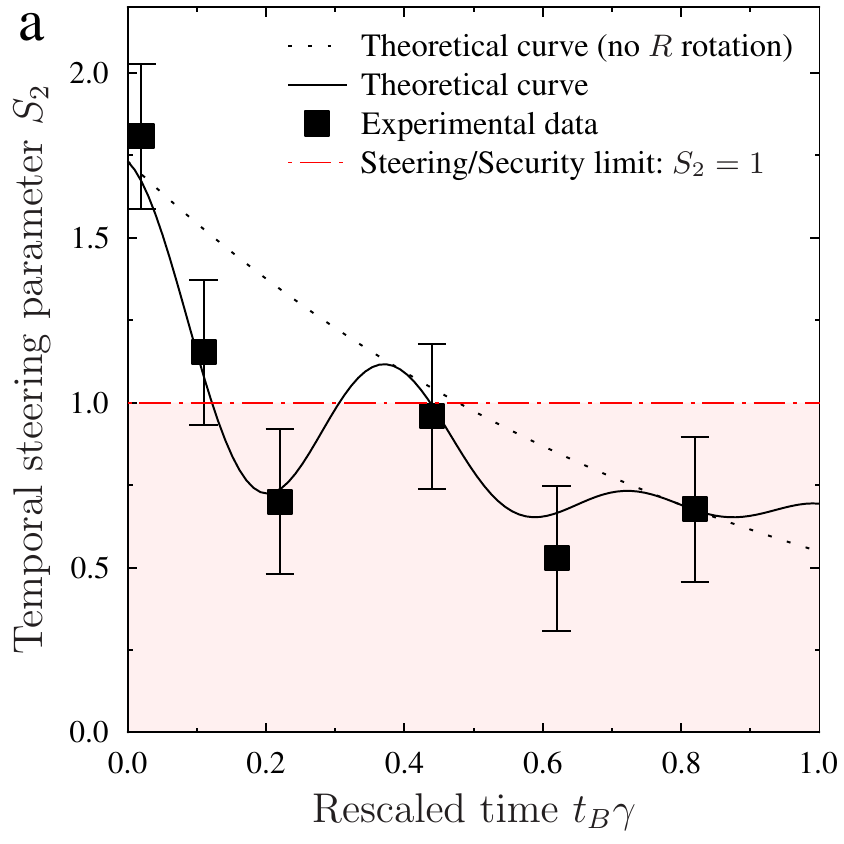}
\includegraphics[width=0.45\linewidth]{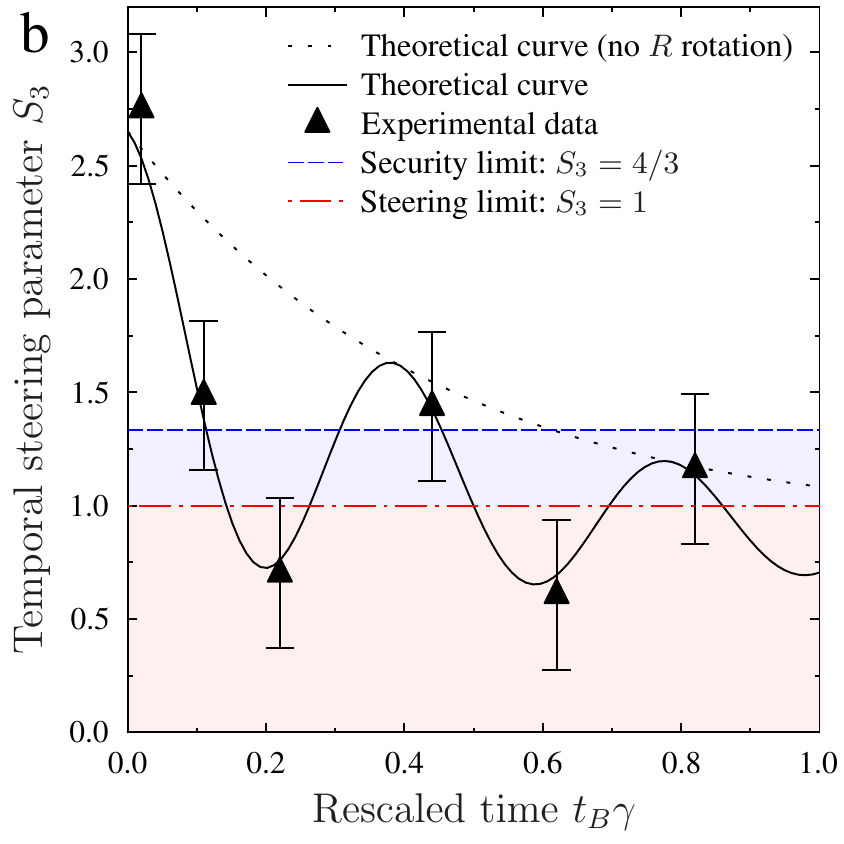}
\caption{Evolution of the TS parameters
$S_N$ corresponding to the TS
inequality~\cite{Chen14,Bart15}: (a) $S_2$ for BB84
(implemented with the eigenvalues of the Pauli operators
$\sigma_1$ and $\sigma_2$) and (b) $S_3$ for B98. Here, $\gamma$
is the damping constant and $t_B$ is the time of the nonunitary
evolution between the measurements of Alice and Bob leading to the
state defined in Eq.~(\ref{eq:state}). The values of $S_3$ and
$S_2$ if the rotation $R(4t_B)$ is not implemented (noise for
$\sigma_1$ and $\sigma_2$ measurements is uniform) are given by
the dotted curves $S_2\approx 2s^{2}\exp(-\gamma t_B)$ and
$S_3\approx s^{2}[2\exp(-2\gamma t_B)+1]$, respectively. This also
corresponds to  our experiment for $4t_B=2n\pi$, where
$n=0,\,1,\,2$. The shrinking factor \( s=0.96\)\ takes into
account the initial impurity of the states sent by Alice. The
solid curves correspond to $S_2$ and $S_3$ calculated for the
state given by Eq.~(\ref{eq:state}) and accounting for the setup
imperfections described in~the Methods.}
\label{fig:2}
\end{figure*}

\begin{figure*}[ht]
\centering
\includegraphics[width=0.45\linewidth]{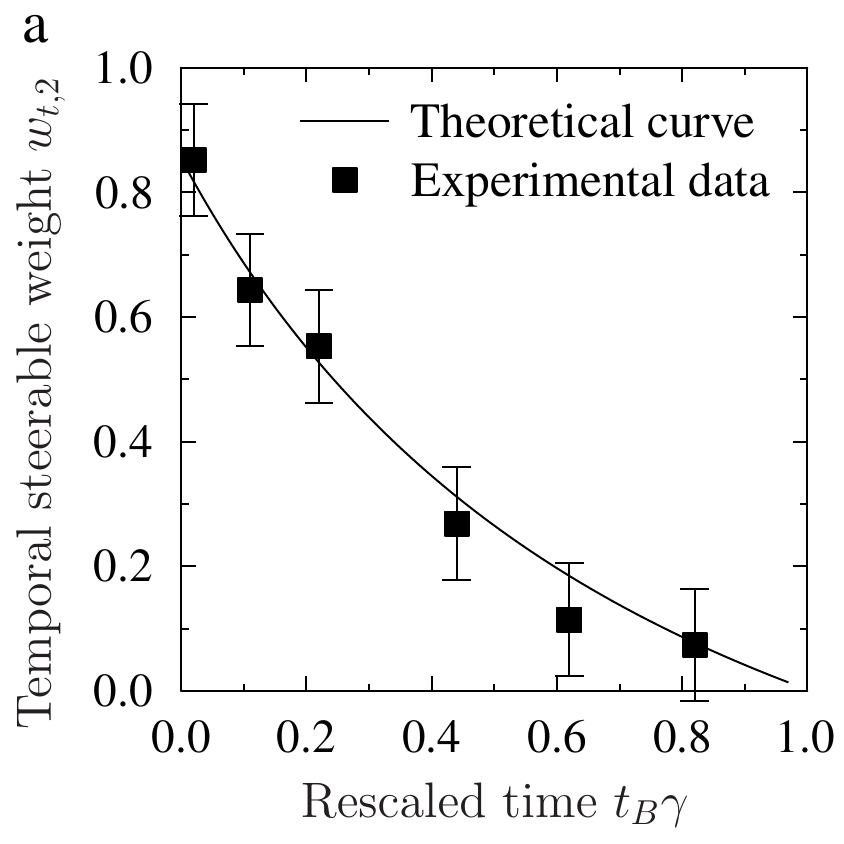}
\includegraphics[width=0.45\linewidth]{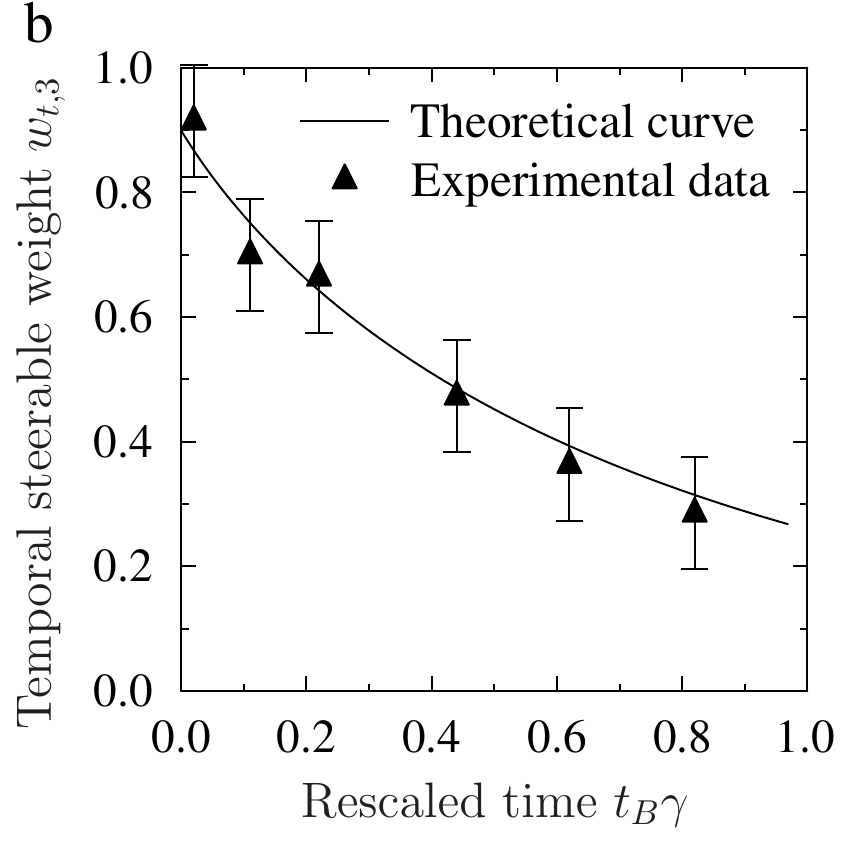}
\caption{Evolution of the temporal steerable
weights $w_{t,N}$ as defined in Eq.~(\ref{eq:sw_def}). Here,
$t_B$ is the nonunitary-evolution time, $\gamma$ is the damping
constant, and $N$ stands for the number of MUB, calculated for the
theoretical (curves) and experimental assemblages (data points)
with a semi-definite program for (a) BB84 ($N=2$) and (b) B98
($N=3$). The solid curves correspond to $w_{t,2}$ and $w_{t,3}$
calculated for the state given by Eq.~(\ref{eq:state}) and
accounting for the setup imperfections described in the
Methods. The temporal steerable
weights do not exhibit an oscillatory behaviour because these do
not depend on the choice of Bob's measurement bases.}
\label{fig:3}
\end{figure*}

\begin{figure*}[ht]
\centering
\includegraphics[width=0.3\linewidth]{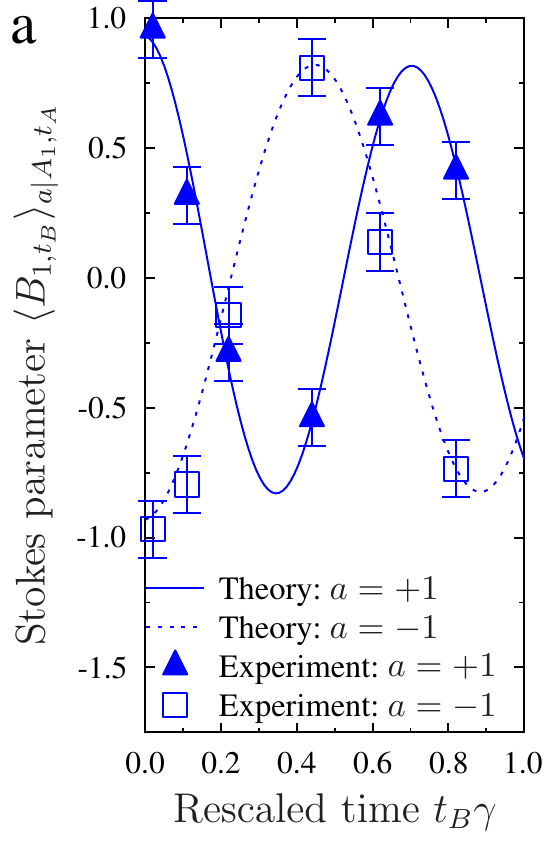}
\includegraphics[width=0.3\linewidth]{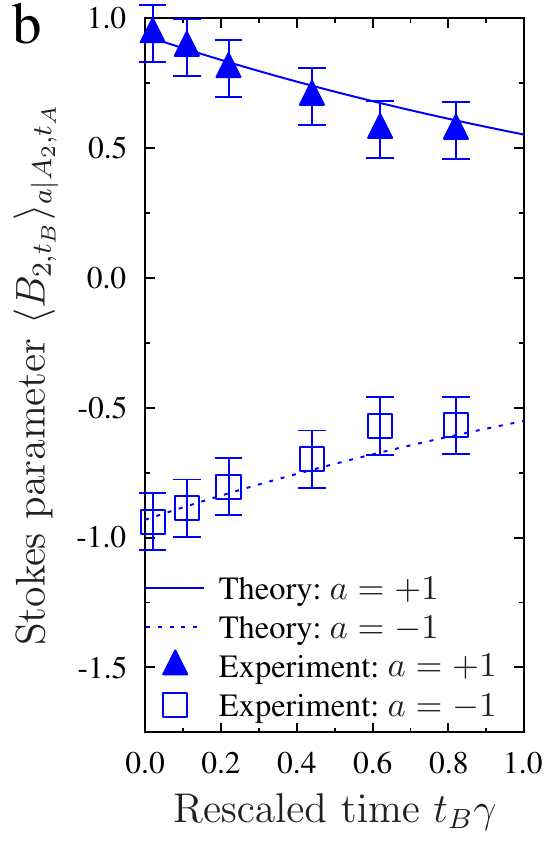}
\includegraphics[width=0.3\linewidth]{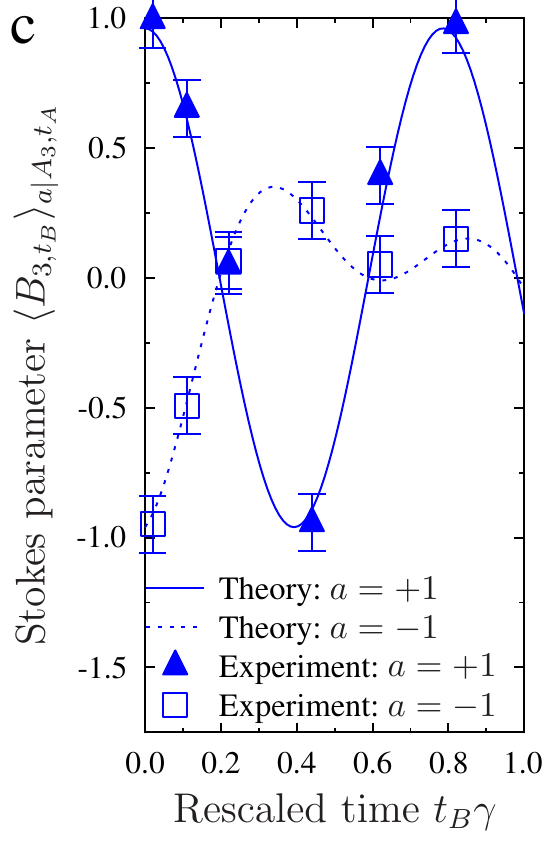}
\caption{Evolution of the generalised Stokes
parameters, \ie, the expected values of the Pauli operators
$B_n=\sigma_n$ (see~Ref.~[\citenum{Bart15}]) for the six eigenstates
$\ket{a,A_n}$ for $n=1,2,3$ and $a=\pm1$. Experimental data for
the relevant states with $a=+1$ ($a=-1$) are illustrated with
triangles (squares). The time between Alice's and Bob's
measurements $t_B$ is given in units of the inverse of the
relaxation constant $\gamma$. In our experiment $\gamma=2\times
10^7\,\mathrm{s}^{-1}$. The solid (dashed) curves correspond to
the expected values of the Pauli operators calculated for the
relevant $a=+1$ ($a=-1$) states defined in Eq.~(\ref{eq:state}).}
\label{fig:4}
\end{figure*}

\end{document}